\documentclass[aps,prl,twocolumn,superscriptaddress,showpacs]{revtex4-1}
\usepackage{graphicx} 

\begin{document}

\title{Density-Dependent Quantum Hall States and Zeeman Splitting in Monolayer and Bilayer WSe$_2$}

\author{Hema C. P. Movva}
\author{Babak Fallahazad}
\author{Kyounghwan Kim}
\author{Stefano Larentis}
\affiliation{Microelectronics Research Center, Department of Electrical and Computer Engineering, The University of Texas at Austin, Austin, Texas 78758, USA}
\author{Takashi Taniguchi}
\author{Kenji Watanabe}
\affiliation{National Institute of Materials Science, 1-1 Namiki Tsukuba, Ibaraki 305-0044, Japan}
\author{Sanjay K. Banerjee}
\author{Emanuel Tutuc}
\email{etutuc@mer.utexas.edu}
\affiliation{Microelectronics Research Center, Department of Electrical and Computer Engineering, The University of Texas at Austin, Austin, Texas 78758, USA}

\date{\today}

\begin{abstract}
We report a study of the quantum Hall states (QHSs) sequence of holes in mono- and bilayer WSe$_2$. The QHSs sequence transitions between predominantly even and predominantly odd filling factors as the hole density is tuned in the range $1.6 - 12\times10^{12}$ cm$^{-2}$. The QHSs sequence is insensitive to the transverse electric field, and tilted magnetic field measurements reveal an insensitivity of the QHSs sequence to the in-plane magnetic field, evincing that the hole spin is locked perpendicular to the WSe$_2$ plane. These observations imply that the QHSs sequence is controlled by the Zeeman-to-cyclotron energy ratio, which remains constant as a function of perpendicular magnetic field at a fixed carrier density, but changes as a function of density due to strong electron-electron interaction.
\end{abstract}


\maketitle


The strong spin-orbit coupling and broken inversion symmetry in $2H$ transition metal dichalcogenide (TMD) monolayers leads to coupled spin and valley degrees of freedom \cite{xiao2012coupled}. Breaking the time reversal symmetry by applying a perpendicular magnetic field further lifts the valley degeneracy, thanks to the spin (valley) Zeeman effect \cite{cai2013magnetic, rose2013spin}. Insights into the Zeeman effect, a fundamental property of TMDs,  have been provided by magneto-optical measurements of TMD monolayers, which report the exciton $g$-factors from luminescence shifts in perpendicular magnetic fields \cite{li2014valley, *aivazian2015magnetic, *macneill2015breaking, *mitioglu2015optical, *wang2015magneto, *stier2016probing, srivastava2015valley}.
While magnetotransport measurements have been traditionally used to determine the effective $g$-factor ($g^*$) in several two-dimensional electron systems (2DESs) \cite{shashkin2001indication, zhu2003spin, vakili2004spin}, the lack of reliable low temperature Ohmic contacts, combined with a moderate mobility, have hampered a similar progress in TMDs. Recent advances in sample fabrication have now facilitated more detailed studies of the electron physics in TMDs \cite{cui2015multi, *wu2016even, fallahazad2016shubnikov, xu2017odd}. Tungsten diselenide (WSe$_2$) is of particular interest because of a large spin-orbit splitting in the valence band \cite{zhu2011giant}, high-mobility \cite{fallahazad2016shubnikov}, and low temperature Ohmic contacts \cite{movva2015high}.

In this work, we report on the magnetotransport of holes in mono- and bilayer WSe$_2$ in the quantum Hall regime. An examination of the Shubnikov--de Haas (SdH) oscillations, and the quantum Hall states (QHSs) sequence reveals interesting carrier density-dependent transitions between predominantly even and predominantly odd filling factors (FFs) as the hole density is tuned. Measurements in tilted magnetic fields reveal an insensitivity of the QHSs sequence to the in-plane magnetic field, indicating that the hole spin is locked perpendicular to the WSe$_2$ plane. The QHSs sequence is also found to be insensitive to the applied transverse electric field. These observations can be explained by a Zeeman-to-cyclotron energy ratio which remains constant as a function of perpendicular magnetic field at a fixed carrier density, but changes as a function of density because of strong electron-electron interaction.

\begin{figure}
\includegraphics[scale=1]{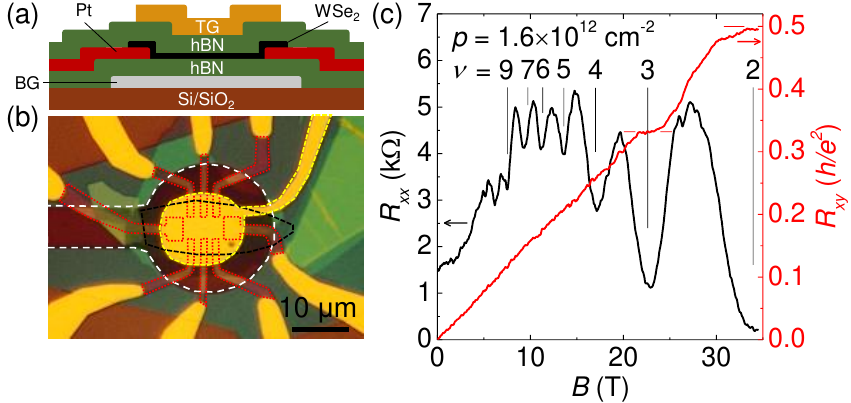}
\caption {\small{(a) Schematic cross section of an hBN encapsulated WSe$_2$ sample with bottom Pt contacts, a top-gate (TG), and a local back-gate (BG). (b) Optical micrograph of a typical WSe$_2$ Hall bar sample. The TG, Pt contacts, WSe$_2$ flake, and BG are outlined in yellow, red, black, and white dashed lines, respectively. (c) $R_{xx}$ and $R_{xy}$ vs $B$ in bilayer WSe$_2$ measured at $T$ = 1.5 K, and at the lowest hole density, $p = 1.6\times10^{12}$ cm$^{-2}$. The $\nu$ values at the $R_{xx}$ minima are labeled. Quantized $R_{xy}$ plateaux are observed at $\nu=2,3$.}}
\label{fig1}
\end{figure}

Figure~\ref{fig1}(a) shows the schematic cross section, and Fig.~\ref{fig1}(b) the optical micrograph of an hBN encapsulated WSe$_2$ sample with bottom Pt contacts, and separate local top- and back-gates.  The mono- and bilayer WSe$_2$ Hall bar samples were fabricated using a modified van der Waals assembly technique \cite{movva2015high, kim2016van}. The bottom Pt electrodes in combination with a large, negative top-gate bias ($V_{TG}$) ensure Ohmic hole contacts to the WSe$_2$ \cite{movva2015high, fallahazad2016shubnikov}. Both $V_{TG}$, and a back-gate bias ($V_{BG}$) were used to tune the WSe$_2$ hole carrier density, $p$. The magnetotransport was probed using low frequency lock-in techniques at a temperature, $T=1.5$ K, and magnetic fields up to $B=35$ T. The $p$ values at which we observe well-defined SdH oscillations are in the range $1.6 - 12\times10^{12}$ cm$^{-2}$, as determined from the slope of the Hall resistance, and from the SdH oscillations minima. The weak interlayer coupling in bilayer WSe$_2$ enables the top layer population with holes, while keeping the bottom layer unpopulated \cite{fallahazad2016shubnikov}. At negative $V_{TG}$, and positive $V_{BG}$, a bilayer effectively acts as a monolayer, albeit with a dissimilar dielectric environment \footnote{While a monolayer has only hBN as the bottom dielectric, a bilayer biased to populate only the top layer has an unpopulated WSe$_2$ bottom layer, which alters the dielectric environment of holes in the top layer.}. All the bilayer data presented here were collected under such biasing conditions, and are therefore closely similar to the monolayer data.

Figure~\ref{fig1}(c) shows the longitudinal ($R_{xx}$) and Hall ($R_{xy}$) resistance vs perpendicular magnetic field ($B$) for a bilayer WSe$_2$ sample at the lowest density, $p=1.6\times10^{12}$ cm$^{-2}$. The $R_{xx}$ data show SdH oscillations starting at $B\cong5$ T, which translates into a mobility, $\mu\simeq2000$ cm$^2/$V~s. The FFs, $\nu=ph/eB$, at the $R_{xx}$ minima are marked. The $R_{xy}$ data show developed QHSs plateaux at $\nu=2,3$, where $R_{xy}$ is quantized at values of $h/{\nu}e^2$; $h$ is the Planck's constant, and $e$ the electron charge. The QHSs occur at consecutive integer FFs ($\nu=2,3,4,...$) for $B>10$ T, indicating a full lifting of the two-fold Landau level (LL) degeneracy in WSe$_2$ \cite{fallahazad2016shubnikov}. For $B<10$ T, the QHSs occur at consecutive odd integer FFs ($\nu=7,9,...$). The occurrence of QHSs at odd FFs contrasts previous observations of QHSs at even FFs in mono- and bilayer WSe$_2$ at higher hole densities \cite{fallahazad2016shubnikov}.  In the following, we will use the term ``QHSs sequence'', be it even or odd, to refer to the QHSs FFs in the lower range of $B$ values, such that the LL degeneracy is not fully lifted.

\begin{figure}
\includegraphics[scale=1]{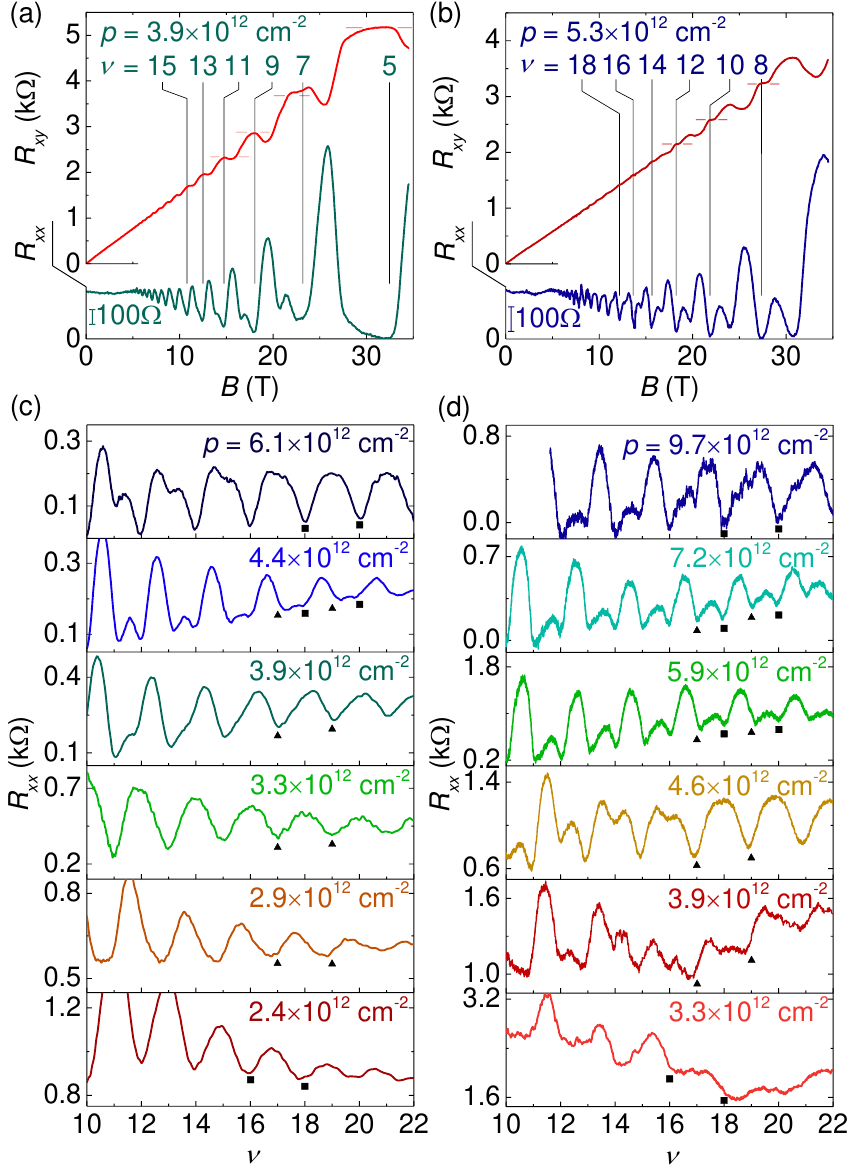}
\caption {\small{(a,b) $R_{xx}$ and $R_{xy}$ vs $B$ in bilayer WSe$_2$ at $p=3.9\times10^{12}$ cm$^{-2}$ [panel (a)], showing QHSs at predominantly odd FFs, and at $p=5.3\times10^{12}$ cm$^{-2}$ [panel (b)], showing QHSs at predominantly even FFs. (c) $R_{xx}$ vs $\nu$ in bilayer WSe$_2$ at different $p$ values. The QHSs sequence changes from even at $p=6.1\times10^{12}$ cm$^{-2}$ to odd at $p=3.3\times10^{12}$ cm$^{-2}$, and back to even at $p=2.4\times10^{12}$ cm$^{-2}$. (d) $R_{xx}$ vs $\nu$ in monolayer WSe$_2$ at different $p$ values show similar QHSs sequence transitions. Representative $R_{xx}$ minima at even and odd QHSs are marked by square and triangle symbols, respectively, in panels (c,d).}}
\label{fig2}
\end{figure}

To better understand the QHSs sequence, we performed magnetotransport measurements as a function of $p$ in both mono- and bilayer WSe$_2$. Figure~\ref{fig2}(a,b) show $R_{xx}$ and $R_{xy}$ vs $B$ measured for the same bilayer sample discussed in Fig.~\ref{fig1} at $p=3.9\times10^{12}$ cm$^{-2}$, and $p=5.3\times10^{12}$ cm$^{-2}$, respectively. While the data at $p=3.9\times10^{12}$ cm$^{-2}$ show an odd QHSs sequence, the QHSs sequence is even at $p=5.3\times10^{12}$ cm$^{-2}$.  Figure~\ref{fig2}(c) shows $R_{xx}$ vs $\nu$ at various values of $p$ from $6.1\times10^{12}$ cm$^{-2}$ to $2.4\times10^{12}$ cm$^{-2}$. The data at $p=6.1\times10^{12}$ cm$^{-2}$ show strong $R_{xx}$ minima at even FFs, and weakly developing minima at odd FFs for $\nu<16$, hence a predominantly even QHSs sequence. As $p$ is reduced to $4.4\times10^{12}$ cm$^{-2}$, the minima at odd FFs become stronger, and equal in strength to the minima at even FFs. The QHSs sequence at this $p$ cannot be unambiguously classified as even or odd. Further reduction of $p$ to $3.9\times10^{12}$ cm$^{-2}$ makes the odd FFs stronger than the even FFs, rendering the QHS sequence as predominantly odd. The odd QHSs sequence is retained down to $p=2.9\times10^{12}$ cm$^{-2}$. On further reduction of $p$ to $2.4\times10^{12}$ cm$^{-2}$, the QHSs sequence reverts to even. Figure~\ref{fig2}(d) shows a similar data set for monolayer WSe$_2$, where the QHSs sequence transitions from even at $p=9.7\times10^{12}$ cm$^{-2}$ to odd at $p=4.6\times10^{12}$ cm$^{-2}$, and back to even at $p=3.3\times10^{12}$ cm$^{-2}$.

This unusual density-dependent QHSs sequence suggests an interesting interplay of the LL Zeeman splitting and the cyclotron energy.  The cyclotron energy of the LLs originating in the upper valence band of monolayer WSe$_2$ is $E_n=-n\hbar\omega_c$; $n$ is the orbital LL index, $\omega_c=eB/m^{*}$ is the cyclotron frequency, $m^{*}=0.45m_0$ the hole effective mass \cite{fallahazad2016shubnikov}; $m_0$ is the bare electron mass. The LLs with $n>0$ are spin-degenerate, whereas the $n=0$ LL is non-degenerate \cite{li2013unconventional, rose2013spin}. Consequently, in the absence of LL Zeeman splitting, an odd QHSs sequence is expected. However, if the LL Zeeman splitting $E_Z=g^{*} \mu_B B$ is comparable to, or larger than the cyclotron energy $E_c=\hbar\omega_c$, the QHSs sequence changes accordingly \cite{chu2014valley}; $\mu_B$ is the Bohr magneton. An $E_Z/{E_c}$ ratio close to an even (odd) integer leads to a QHSs sequence that is predominantly odd (even). Two noteworthy observations can be made based on Fig.~\ref{fig2} data.  First, the presence of an even or odd QHSs sequence at a fixed density implies that the $E_Z/{E_c}$ ratio, and $g^{*}$ do not change with $B$ at low fields. Second, the different QHSs transitions observed in Fig.~\ref{fig2} suggest that the $E_Z/{E_c}$ ratio, and therefore $g^{*}$ change with density, likely because electron-electron interaction associated with the large $m^{*}$ in this system leads to an enhanced $g^*$ as the density is reduced.

\begin{figure}
\includegraphics[scale=1]{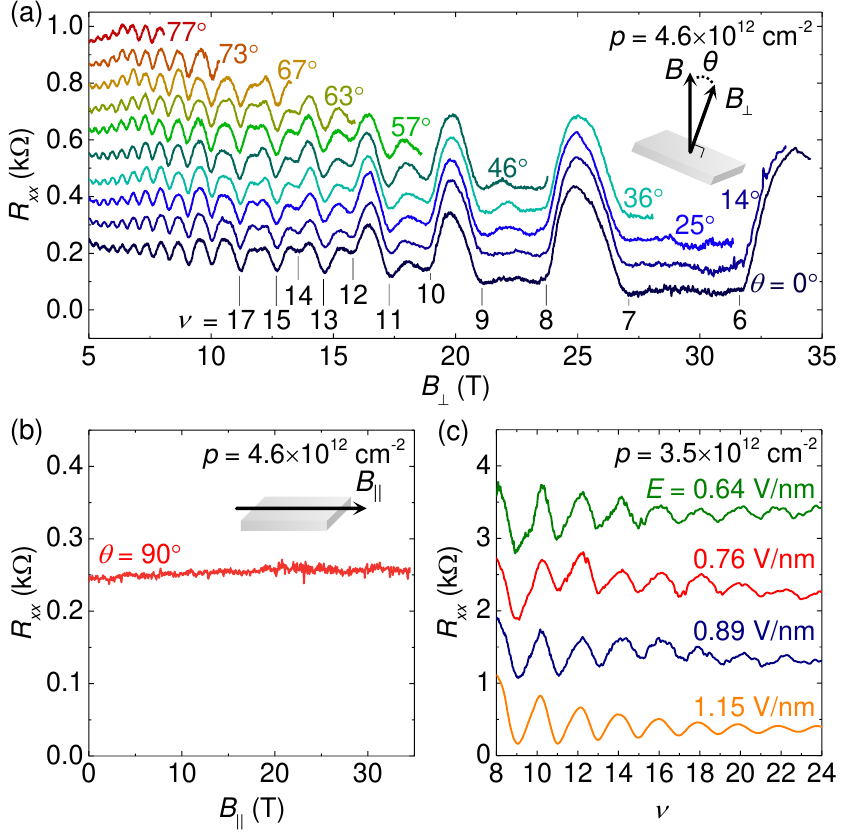}
\caption {\small{(a) $R_{xx}$ vs $B_\perp$ in monolayer WSe$_2$ at $p=4.6\times10^{12}$ cm$^{-2}$, and at different $\theta$ values. The traces are offset for clarity. The QHSs sequence shows no change with $\theta$. Inset: schematic of the sample orientation with respect to the $B$-field. (b) $R_{xx}$ vs $B_{||}$ corresponding to the $\theta=90^{\circ}$ trace of panel (a) data. The $R_{xx}$ remains unchanged in the entire $B$-field range. (c) $R_{xx}$ vs $\nu$  measured in bilayer WSe$_2$ at $p=3.5\times10^{12}$ cm$^{-2}$, and at different $E$-field values.  The data show that the QHSs sequence does not change with the $E$-field. The traces are offset for clarity.}}
\label{fig3}
\end{figure}

Two measurement types have been traditionally used to probe the Zeeman splitting in 2DESs. In a tilted magnetic field, the $B$ component perpendicular to the 2DES plane ($B_\perp$) determines the cyclotron energy $E_c=\hbar\omega_c=\hbar{eB_\perp}/m^{*}$, while the Zeeman energy, $E_Z=g^{*}\mu_B{B}$ depends on the total field \cite{fang1968effects}. At specific angles $\theta$ between the $B$-field and the normal to the 2DES plane, the $E_Z/E_c$ ratio attains integer values, which leads to a collapse of different QHSs, and allows a quantitative determination of $E_Z$. To assess this effect in our samples, Fig.~\ref{fig3}(a) shows $R_{xx}$ vs $B_\perp$ for a monolayer WSe$_2$ sample, measured at $p=4.6\times10^{12}$ cm$^{-2}$, and at different values of $\theta$. The $R_{xx}$ at $\theta=0^\circ$ shows an odd QHSs sequence, which remains virtually unchanged for all values of $\theta$ up to $77^\circ$. A similar behavior was observed even for bilayer WSe$_2$, suggesting indeed that $E_Z$ is {\it insensitive} to $\theta$, and in turn, to the parallel component of the $B$-field ($B_{||}$) in both mono- and bilayer WSe$_2$. This observation is in stark contrast to the vast majority of 2DESs explored in host semiconductors such as Si \cite{fang1968effects, shashkin2001indication}, GaAs \cite{zhu2003spin}, AlAs \cite{vakili2004spin}, black phosphorus \cite{li2016quantum}, and bulk WSe$_2$ \cite{xu2017odd}.

A second technique used to determine $E_Z$ is the magnetoresistance measured as a function of the magnetic field parallel to the 2DES plane. The Zeeman coupling leads to a spin polarization of the 2DES, which reaches unity when $E_Z$ is equal to the Fermi energy. Experimentally, $R_{xx}$ vs $B_{||}$ measured at $\theta=90^\circ$ shows a positive magnetoresistance, along with a saturation or a marked kink at the $B$-field corresponding to full spin polarization \cite{okamoto1999spin, tutuc2001plane, zhu2003spin, vakili2004spin}. Figure~\ref{fig3}(b) shows $R_{xx}$ vs $B_{||}$ data for the monolayer sample of Fig.~\ref{fig3}(a). Surprisingly, yet consistent with Fig.~\ref{fig3}(a) data, $R_{xx}$ remains constant over the entire range of $B_{||}$, which implies that $E_Z$ depends only on $B_{\perp}$, namely $E_Z=g^* \mu_B B_{\perp}$, via a density-dependent $g^*$. The insensitivity of $E_Z$ to $B_{||}$ indicates that the hole spin is locked perpendicular to the plane, a direct consequence of the strong spin-orbit coupling, and mirror symmetry in monolayer WSe$_2$ \cite{zhu2011giant}. Optical experiments on monolayer WSe$_2$ have shown a similar insensitivity of $E_Z$ to $B_{||}$ \cite{srivastava2015valley}. We note that spin-locking along the $z$-direction renders the tilted $B$-field technique ineffective to determine $E_Z$.

\begin{figure*}
\includegraphics[scale=1]{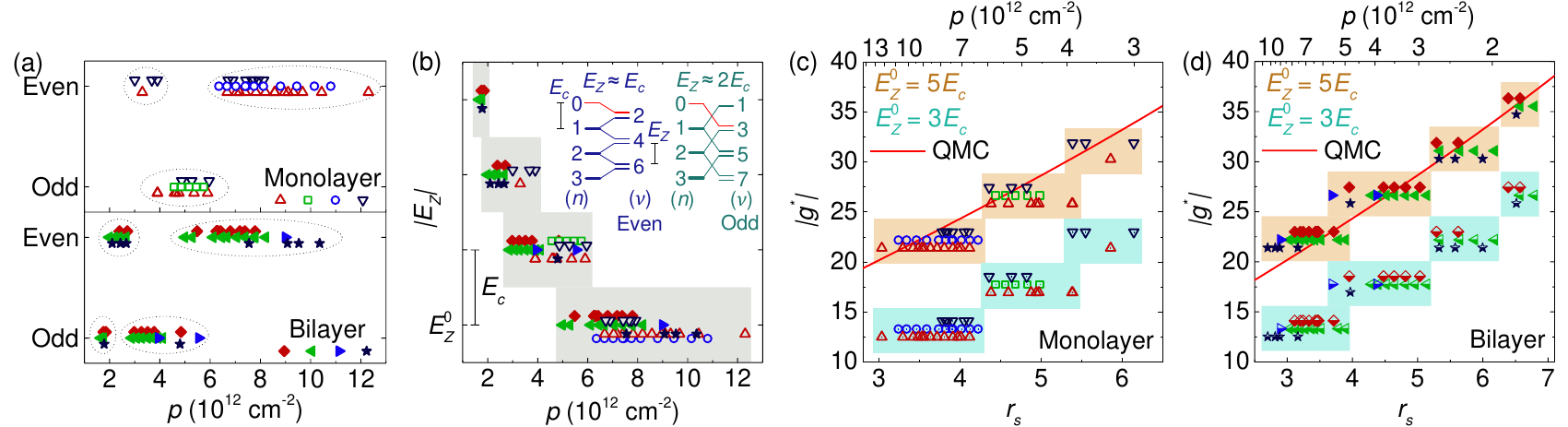}
\caption {\small{(a) QHSs sequence vs $p$ for four monolayer (top, open symbols), and four bilayer (bottom, solid symbols) WSe$_2$ samples. The dotted lines group the data points belonging to the same QHSs sequence in a given $p$ range. (b) Panel (a) data converted to $|E_Z|$ vs $p$, assuming an $|E_Z|$ increment of $E_c$ for every QHSs sequence transition in the direction of decreasing $p$. The inset shows two possible scenarios where the QHSs sequence could be even ($E_Z\approx{E_c}$) or odd ($E_Z\approx{2E_c}$). (c) Monolayer, and (d) Bilayer WSe$_2$ $|g^*|$ vs $r_s$ (bottom axis) or $p$ (top axis) for $E_Z^0=5E_c$ (open, solid symbols), $E_Z^0=3E_c$ (dotted, half-filled symbols), and the QMC calculation for $g_b=8.5$ (line). The symbols within a group are vertically offset for clarity. The shaded regions correspond to a $\pm{E_c/2}$ error bar in panel (b) and a $\pm{\Delta{g^*}}/2$ error bar in panels (c,d).}}
\label{fig4}
\end{figure*}

In light of Fig.~\ref{fig2} data which suggest a density-dependent $g^{*}$, one important question is whether the $g^{*}$ variation is determined by the density, or by the applied transverse electric field ($E$), which depends on the applied gate biases and can change concomitantly with the density. The impact of a transverse $E$-field on bandstructure has been experimentally investigated, among others, in 2D electrons in InGaAs/InAlAs \cite{nitta1997gate}, 2D holes in GaAs \cite{lu1998tunable, *winkler2005anomalous}, and has been theoretically considered in TMDs using a Bychkov-Rashba coupling \cite{kormanyos2014spin}. To probe the impact of the $E$-field on the QHSs sequence in WSe$_2$, we performed $R_{xx}$ vs $B$ measurements by varying $E=|C_{TG}V_{TG}-C_{BG}V_{BG}|/2\epsilon_0$ at constant $p$; $C_{TG}$ ($C_{BG}$) is the top (back)-gate capacitance, and $\epsilon_0$ the vacuum permittivity. Figure~\ref{fig3}(c) shows $R_{xx}$ vs $\nu$ measured in bilayer WSe$_2$ at $p=3.5\times10^{12}$ cm$^{-2}$, at different values of $E$. The data show no variation of the QHSs sequence when the $E$-field varies from 0.64 V/nm to 1.15 V/nm. By comparison, the $E$-field changes from 0.92 V/nm to 1.11 V/nm in Fig.~\ref{fig2}(c), concomitantly with the density change from $6.1\times10^{12}$ cm$^{-2}$ to $3.9\times10^{12}$ cm$^{-2}$, a range in which a QHSs sequence transition from even to odd is observed. Based on these observations, we rule out the effect of the $E$-field on $g^*$, and in turn, on the QHSs sequence.

In Fig.~\ref{fig4}(a), we summarize the QHSs sequence vs $p$ for four monolayer, and four bilayer WSe$_2$ samples. The data points are grouped into an even or odd QHSs sequence over a range of $p$. We attribute the QHSs sequence transitions to a change in the $E_Z/E_c$ ratio with varying $p$. For instance, $E_Z\approx{E_c}$ ($E_Z\approx{2E_c}$) can lead to an even (odd) QHSs sequence [Fig.~\ref{fig4}(b) inset]. Generalized further, $|E_Z|/E_c\in[2k-1/2,2k+1/2]$ yields an odd QHSs sequence, and $|E_Z|/E_c\in[2k+1/2,2k+3/2]$ yields an even QHSs sequence; $k$ is an integer. Each of the groups of Fig.~\ref{fig4}(a) can therefore be assigned an $|E_Z|$ within a $[-E_c/2,E_c/2]$ window. Starting with $|E_Z|=E_Z^0$ at the highest value of $p$ probed, and assuming $|E_Z|$ increases with reducing $p$ because of interaction, we can assign an $|E_Z|$ increment of $E_c$ for every QHSs sequence transition in the direction of decreasing $p$ [Fig.~\ref{fig4}(b)]. In the absence of electron-electron interaction, the $g$-factor, referred to as the band $g$-factor ($g_b$) is determined by the material bandstructure. Exchange interaction can enhance $g_b$ to a value $g^*$, which increases with decreasing density, an observation reported for several 2DESs in Si \cite{okamoto1999spin, shashkin2001indication}, GaAs \cite{zhu2003spin}, and AlAs \cite{vakili2004spin}. The interaction strength is gauged by the dimensionless parameter, $r_s=1/(\sqrt{\pi{p}}a^*_B)$, the ratio of the Coulomb energy to the kinetic energy; $a^*_B=a_B(\kappa{m_0}/m^*)$, $a_B$ is the Bohr radius, and $\kappa$ is the effective dielectric constant of the medium surrounding the 2DES.

The $|E_Z|$ vs $p$ of Fig.~\ref{fig4}(b) can therefore be converted to a $|g^*|$ vs $r_s$ dependence. We first address the value of $E_Z^0=g^*_0\mu_BB$. The even QHSs sequence at the highest $p$ probed implies that $E_Z^0=(2k+1)E_c$, or equivalently, $g^*_0=4.44(2k+1)$; $k$ is an integer \footnote{We assume a density independent $m^*=0.45m_0$ \cite{fallahazad2016shubnikov}}. Recent magneto-reflectance measurements that resolve the LL spectrum report a $g_b=8.5$ for holes in monolayer WSe$_2$ \cite{wang2016valley}. To account for the uncertainty in $E_Z^0$, we consider two scenarios of $g^*_0$ corresponding to $k=1$ ($E_Z^0=3E_c$), and $k=2$ ($E_Z^0=5E_c$). We rule out the case $k=0$ based on the reported $g_b$ value \cite{wang2016valley}. The $E_c$-step increments of $|E_Z|$ between groups are equivalent to a $|g^*|$ increment of $\Delta{g^*}=2m_0/m^*=4.44$ \cite{Note2}. Within this framework, Fig.~\ref{fig4}(c) and Fig.~\ref{fig4}(d) show $|g^*|$ vs $r_s$ for the mono- and bilayer samples, respectively.  Because of the difference in dielectric environment, slightly different $\kappa$ values were used to convert $p$ into $r_s$ for mono- and bilayer WSe$_2$ \footnote{For a 2DES with an asymmetric dielectric environment, $\kappa(t,b)=[(\epsilon^{||}_t\epsilon^{\perp}_t)^{1/2}+(\epsilon^{||}_b\epsilon^{\perp}_b)^{1/2}]/2$; $\epsilon^{||/\perp}_{t(b)}$ are the top (bottom) medium dielectric constants. For monolayer WSe$_2$, $\kappa=\kappa(hBN,hBN)$. For bilayer WSe$_2$ \cite{Note1}, $\kappa=[\kappa(hBN,hBN)+\kappa(hBN,WSe_2)]/2$. We use $\epsilon_{hBN}^{||}=3.0$, $\epsilon_{hBN}^{\perp}=6.9$ \cite{geick1966normal}, $\epsilon_{WSe_2}^{||}=7.2$ \cite{kim2015band}, and $\epsilon_{WSe_2}^{\perp}=14$ \cite{li2014measurement}.}.

For comparison, in Fig.~\ref{fig4}(c,d) we include the $g_b$ value multiplied by the interaction enhanced spin susceptibility obtained from quantum Monte Carlo (QMC) calculations \cite{attaccalite2002correlation}. The QMC calculations along with the $g_b$ of Ref.~\cite{wang2016valley} match well with $|g^*|$ determined using $E_Z^0=5E_c$ for both mono- and bilayer WSe$_2$. Noteworthy, the relatively large $m^*=0.45m_0$ leads to moderately large $r_s$ values, and potentially strong interaction effects even at high carrier densities \cite{tanatar1989ground}.

In summary, we present a density-dependent QHSs sequence of holes in mono- and bilayer WSe$_2$, which transitions between even and odd filling factors as the hole density is tuned. The QHSs sequence is insensitive to the in-plane $B$-field, indicating that the hole spin is locked perpendicular to the WSe$_2$ plane, and is also insensitive to the transverse $E$-field. The QHSs sequence transitions stem from an interplay between the cyclotron and Zeeman splittings via an enhanced $g^*$ due to strong electron-electron interaction.

\begin{acknowledgments}
We thank X. Li, K. F. Mak, and F. Zhang for technical discussions. We also express gratitude to D. Graf, J. Jaroszynski, and A. V. Suslov for technical assistance. We acknowledge support from NRI SWAN, National Science Foundation Grant No. EECS-1610008, and Intel Corp. A portion of this work was performed at the National High Magnetic Field Laboratory, which is supported by National Science Foundation Cooperative Agreement No. DMR-1157490, and the State of Florida.

H. C. P. M. and B. F. contributed equally to this study.
\end{acknowledgments}

%

\end{document}